\documentstyle[11pt,newpasp]{article}

\begin{document}

\title{Single CO peak in the double bar galaxy NGC 5850}

\author{S. Leon}
\affil{Institute of Astronomy \& Astrophysics, Academia Sinica, Nankang,
Taipei, Taiwan 115}

\author{F. Combes}
\affil{DEMIRM, Observatoire de Paris, Paris, F-75014}

\author{D. Friedli}
\affil{Observatoire de Gen\`eve, Sauverny, CH-1290}

\section{Introduction}

NGC 5850 is the prototype of double barred galaxy (Friedli et al., 1996) classified
as SBb(sr) I-II (Higdon et al. 1998, Prieto et al. 1997). This kind of system is 
primordial to understand the physical mechanism responsible for feeding 
galaxy nuclei and boost the star formation rate. Schlosman et al. (1989) proposed 
 that the nuclear bar would produce the inwards inflow of the molecular
gas trapped on the primary ILR through the nuclear bar resonances.
Higdon et al. (1998) emphasized that NGC~5850 has likely 
been perturbed by a high-speed
encounter with the nearby massive elliptical NGC~5846.

\section{Molecular gas and other tracers}
We have observed the CO(1-0) emission of the galaxy NGC 5850: mapped
the very center, using the IRAM Plateau de Bure
interferometer, to reach a 2.4''$\times$1.5'' (PA=-165$^\circ$) spatial resolution 
(Fig. \ref{fig_co_imj}),  and mapped the primary bar with the IRAM-30m telescope,
with a $22''$ beam. We estimate the flux missed
by the interferometer to about 40 \% of the single-dish flux. The total flux in the center
is 41.7 Jy/beam for the interferometer.
To estimate the H$_2$ surface density in the center, 
we used the standard conversion factor
N(H$_2$)/I$_{CO}$=2.3 10$^{20}$ cm$^{-2}$K$^{-1}$km$^{-1}$s. 
The northern concentration of gas, 
reaches a surface density of 200 M$_\odot$pc$^{-2}$ (not including 30\% He). 
The total H$_2$ towards the center is 
about 6.7 10$^7~M_\odot$ with the interferometer. Using the single-dish, 
we find that the primary bar has about 3.4 10$^9$ M$_\odot$ of 
molecular gas.


\begin{figure*}
\includegraphics{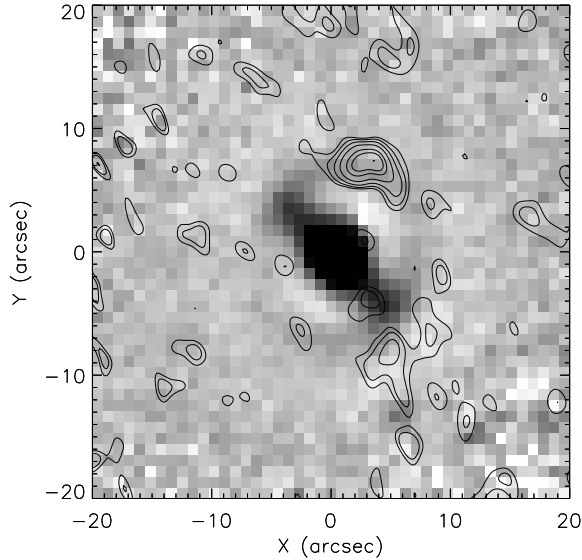}
\vspace{6cm}
\caption{ Small scale details (spatial filtering scale $< 1.6''$) on
the J image overlaid by CO(1$\to$0) integrated intensity contours
(0.01, 0.012, 0.016, 0.02, 0.025, 0.03, 0.04, 0.05, 0.063, 0.08, 0.1 Jy/beam).}
\end{figure*}

The critical surface density $\Sigma_c$ for gravitational instabilities (e.g. 
Kennicutt 1989) is:
$\Sigma_c=\alpha\frac{\kappa\sigma}{3.36G}$, with $\kappa$ the epicyclic frequency,
$\sigma$ the velocity dispersion of the gas and $\alpha$ is a constant
close to unity. 
The northern concentration, likely made of a collection of Giant Molecular
Clouds (GMC), has a {\em global} velocity dispersion of $\sim$ 75 km/s.
The critical surface density is then
1350 M$_\odot$pc$^{-2}$ which is much higher than the peak value observed. 
It may explain why no star formation has occured in that great reservoir of molecular gas 
as traced by the H$_\alpha$ emission. The HI gas (3.3 10$^9$ M$_\odot$) is more 
concentrated in the larger inner ring and outer arms of the galaxy (Higdon et al. 1998).


\section{Conclusion}
We have found CO emission in the center of NGC~5850, located in a single peak
on the northern part of the nuclear ring. 
The high velocity dispersion of the molecular gas may prevent
star formation in that region. The CO distribution in the center of barred 
galaxies is generally found to be either in the nucleus 
(Garc\'ia-Burillo et al. 1998) or trapped in twin peaks related to the 
resonances of the bars (Kenney et al. 1992,
Downes et al. 1996, Garc\'ia-Burillo et al. 1998). Gas simulations performed with a 
single bar pattern and without the tidal influence of the companion NGC 5846 are 
unable to reproduce the features observed in NGC~5850 (Combes, Leon \& Buta 1998).
The decoupling of a second bar appears necessary. The presence of
the single molecular peak could be due to an $m=1$ mode excited by the massive companion.

\small{

}

\end{document}